\documentclass[epj]{svjour}
\usepackage{graphics}
\usepackage{cite}
\usepackage{graphicx}
\usepackage{amsmath,amssymb}
\usepackage{color}

\begin{document}
\title{Experimental and numerical investigations of switching wave dynamics in a normally dispersive fiber ring resonator}
\author{Bruno Garbin\inst{1} \and Yadong Wang\inst{1} \and Stuart G. Murdoch\inst{1} \and Gian-Luca Oppo\inst{2} \and St\'ephane Coen\inst{1} \and Miro Erkintalo\inst{1}}

\titlerunning{Experimental and numerical investigations of switching wave dynamics \ldots}
\authorrunning{Garbin et al}

\institute{The Dodd-Walls Centre for Photonic and Quantum Technologies, Department of Physics, The University of Auckland, Auckland 1142, New Zealand \and SUPA and Department of Physics, University of Strathclyde, Glasgow G4 0NG, Scotland, EU}
\date{Received: date / Revised version: date}

\abstract{Optical frequency combs generated in normally dispersive Kerr microresonators have been observed to correspond to dark temporal structures, and theoretically explained as interlocked switching waves (also known as domain walls or fronts). The time-domain dynamics that underpin the formation of this type of frequency combs has however so far eluded direct experimental observation. Here we use a closely related system --- a synchronously driven optical fibre ring resonator --- to experimentally study the dynamics of deterministically excited switching waves. We measure the switching wave velocities across broad parameter regions, and observe clear signatures of interlocking behaviour leading to the formation of persisting dark pulses. Our experimental findings are in good agreement with simulations of the mean-field Lugiato-Lefever equation, and strongly support the nature of normal-dispersion microresonator frequency combs suggested in the literature.
} %end of abstract
\maketitle
\section{Introduction}
\label{intro}

Coherently-driven dispersive Kerr resonators have attracted considerable interest over the last ten years. On the one hand, macroscopic ring cavities constructed from standard single-mode optical fibres~\cite{haelterman_dissipative_1992, coen_experimental_1998} have been used for fundamental studies of dissipative patterns and localized structures known as temporal cavity solitons (TCSs), which can be applied as bits in all optical buffers~\cite{coen_continuous-wave_2001, leo_temporal_2010, leo_dynamics_2013, jang_ultraweak_2013, jang_temporal_2015, copie_competing_2016, jang_all-optical_2016}. On the other hand, monolithic microresonators have enabled the generation of optical frequency combs~\cite{delhaye_optical_2007, kippenberg_microresonator-based_2011, savchenkov_kerr_2011, ferdous_spectral_2011,herr_temporal_2014, henriet_kerr_2015, yi_soliton_2015, joshi_thermally_2016, webb_experimental_2016} whose unique and attractive characteristics make them ideally suited for applications such as telecommunications~\cite{pfeifle_coherent_2014, pfeifle_optimally_2015} and spectroscopy~\cite{suh_microresonator_2016, dutt_-chip_2016arXiv}.

Besides scaling factors associated with the differences in physical size, microresonators and fibre ring resonators display very similar dynamics~\cite{leo_temporal_2010, coen_modeling_2013, coen_universal_2013}. In fact, both systems can be described by the very same mean-field equation~\cite{coen_modeling_2013, chembo_spatiotemporal_2013} that is analogous to the celebrated Lugiato-Lefever equation (LLE) of spatially diffractive cavities~\cite{lugiato_spatial_1987}. This similarity has, in particular, allowed fibre ring resonator experiments to provide invaluable insights into comb dynamics that cannot be resolved in microresonators due to their small size~\cite{luo_spontaneous_2015, jang_controlled_2016, anderson_observations_2016}. So far, the vast majority of studies have focused on resonators that exhibit anomalous dispersion (at the driving wavelength), arguably because the modulational instabilities (MI) that underpin the formation of cavity solitons and optical frequency combs manifest themselves over a much wider (and in general more accessible) range of parameters compared to resonators exhibiting normal dispersion~\cite{coen_modulational_1997, coen_competition_1999}. The difficulty of achieving anomalous dispersion in certain key regions of the electromagnetic spectrum has, however, also prompted research into better understanding the dynamics and opportunities in the normal dispersion regime, especially in the context of microresonator frequency combs~\cite{matsko_normal_2012, godey_stability_2014, coillet_azimuthal_2013, liang_generation_2014, xue_mode-locked_2015, xue_second-harmonic_2016arXiv}. Of particular significance has been the realization that higher-order effects can considerably expand the range of parameters over which MI can occur, thus rendering the creation of frequency combs feasible in the normal dispersion regime~\cite{tlidi_control_2007, copie_competing_2016, xue_mode-locked_2015, xue_second-harmonic_2016arXiv}. Indeed, Xue et al have convincingly demonstrated that, normal dispersion notwithstanding, interactions between different mode families can alter the phase of spectral components around the driving wavelength, permitting phase-matching of the four-wave mixing that underlies both MI and the subsequent formation of broadband frequency combs~\cite{xue_mode-locked_2015, xue_second-harmonic_2016arXiv}.

A remarkable finding reported in~\cite{xue_mode-locked_2015, xue_second-harmonic_2016arXiv} is that, following the initial MI activated by higher-order effects, the intracavity field may evolve into a fully phase-locked state corresponding to dark pulses --- localized low-intensity dips --- embedded in a high-intensity continuous-wave (cw) background. Several theoretical models have been put forward to explain the physical essence of these dark structures~\cite{xue_mode-locked_2015, lobanov_frequency_2015, parra-rivas_origin_2016}. Parra-Rivas et al made the particularly convincing case that the dark structures consist of two interlocked switching waves (SWs)~\cite{parra-rivas_origin_2016}. SWs, also known in the literature as domain walls or fronts, are traveling front solutions connecting the two homogenous states of a bistable system~\cite{rozanov_transverse_1982}. They have been extensively investigated in many different contexts such as chemical reactions~\cite{petrov_resonant_1997}, biological~\cite{dawson_fire-diffuse-fire_1999} and magnetic systems~\cite{eschenfelder_magnetic_1981}, granular media~\cite{macias_spatially_2013}, hydrodynamics~\cite{pomeau_front_1986}, population dynamics~\cite{clerc_patterns_2005}, flame front propagation~\cite{williams_combustion_1965}, as well as nonlinear optics~\cite{rozanov_transverse_1982, coen_convection_1999, oppo_domain_1999, gomila_stable_2001, residori_patterns_2005}.

Under purely diffusive transverse coupling, SWs have hyperbolic tangent profiles and they travel until the more stable state has fully invaded the less stable one~\cite{firth_diffusive_1985}. There exists only one particular set of parameters --- known as the Maxwell point --- at which the fronts remain stationary. In the presence of dispersion or diffraction, however, SWs can display oscillatory tails~\cite{ackemann_chapter_2009}, which allow two travelling SWs to interact and trap each other (akin to the formation of soliton bound states~\cite{malomed_bound_1993, schapers_interaction_2000}). This results in the arrest of their relative motion, thereby giving rise to a stable localized region of one state embedded in the other~\cite{rosanov_diffractive_1990, oppo_domain_1999, gallego_self-similar_2000, oppo_characterization_2001, pesch_two-dimensional_2007, ackemann_chapter_2009}. Before recently attracting attention in the context of frequency comb generation in \emph{dispersive} nonlinear optical resonators, such front locking dynamics have been extensively studied in spatially \emph{diffractive} systems with one and two transverse dimensions~\cite{rozanov_transverse_1982, rosanov_diffractive_1990, oppo_domain_1999}.

As demonstrated by Parra-Rivas et al (based on theoretical analysis of the LLE), interlocked SWs can qualitatively explain many of the salient features observed in experiments involving frequency comb generation in normally dispersive microresonators~\cite{parra-rivas_origin_2016}. However, the small physical size of microresonators obstructs quantitative study of the transient SW dynamics and their behaviour at the point of interlocking. Macroscopic fibre ring resonators are devoid of this obstruction. They should therefore permit more controlled investigations but so far only steady-state SW observations have been reported~\cite{coen_convection_1999}.

In this Article, we experimentally and numerically study the transient dynamics of deterministically excited time-domain SWs and their interlocking. Specifically, we use a synchronously-driven, normally dispersive fibre ring resonator, and demonstrate robust SW excitation by intensity modulation of the flat-top nanosecond pulses driving the resonator. By monitoring the transient SW dynamics in real time, we are able to experimentally measure the SW velocity across a wide range of parameters, as well as the transient build-up of interlocked states. Our experiments are in good agreement with simulations of the LLE, and they strongly corroborate the proposition that mode-locked frequency combs in normally dispersive microresonators correspond to interlocked SWs.

\section{Illustrative numerical simulations}

We begin by briefly recalling the physics of SWs and their interlocking behaviour. To this end, we present results from numerical simulations pertinent to the experiments that will follow.

It is well known that, in the limit of high cavity finesse, the dynamics of a coherently driven fibre ring resonator is governed by a mean field equation that is formally identical to the celebrated Lugiato-Lefever equation of spatially diffractive Kerr cavities~\cite{haelterman_dissipative_1992, lugiato_spatial_1987}. Adopting the same normalization as in~\cite{leo_temporal_2010}, the equation reads:
\begin{equation}
 \frac{\partial E (t,\tau)}{\partial t} = \left[-1 + i(|E|^2 - \Delta) - i \eta \frac{\partial^2}{\partial \tau^2} \right] E + \sqrt{X}.
\label{LLE}
\end{equation}
Here the variable $t$ corresponds to a ``slow'' time that describes the evolution of the electric field envelope $E(t,\tau)$ inside the resonator at the scale of the cavity photon lifetime, while $\tau$ corresponds to a ``fast'' time defined in a reference frame moving at the group-velocity of light at the driving wavelength and describes the temporal profile of the intracavity field envelope. The terms on the right-hand side of Eq.~\eqref{LLE} describe --- from left to right --- the cavity losses, the Kerr nonlinearity, the detuning between the driving frequency and the closest cavity resonance ($\Delta$ is the detuning parameter), second-order group-velocity dispersion ($\eta$ is the sign of the group-velocity dispersion coefficient $\beta_2$), and the coherent driving of the cavity ($X$~represents the driving power).

\begin{figure}[b]
  \centerline{\includegraphics[width=1.0 \columnwidth]{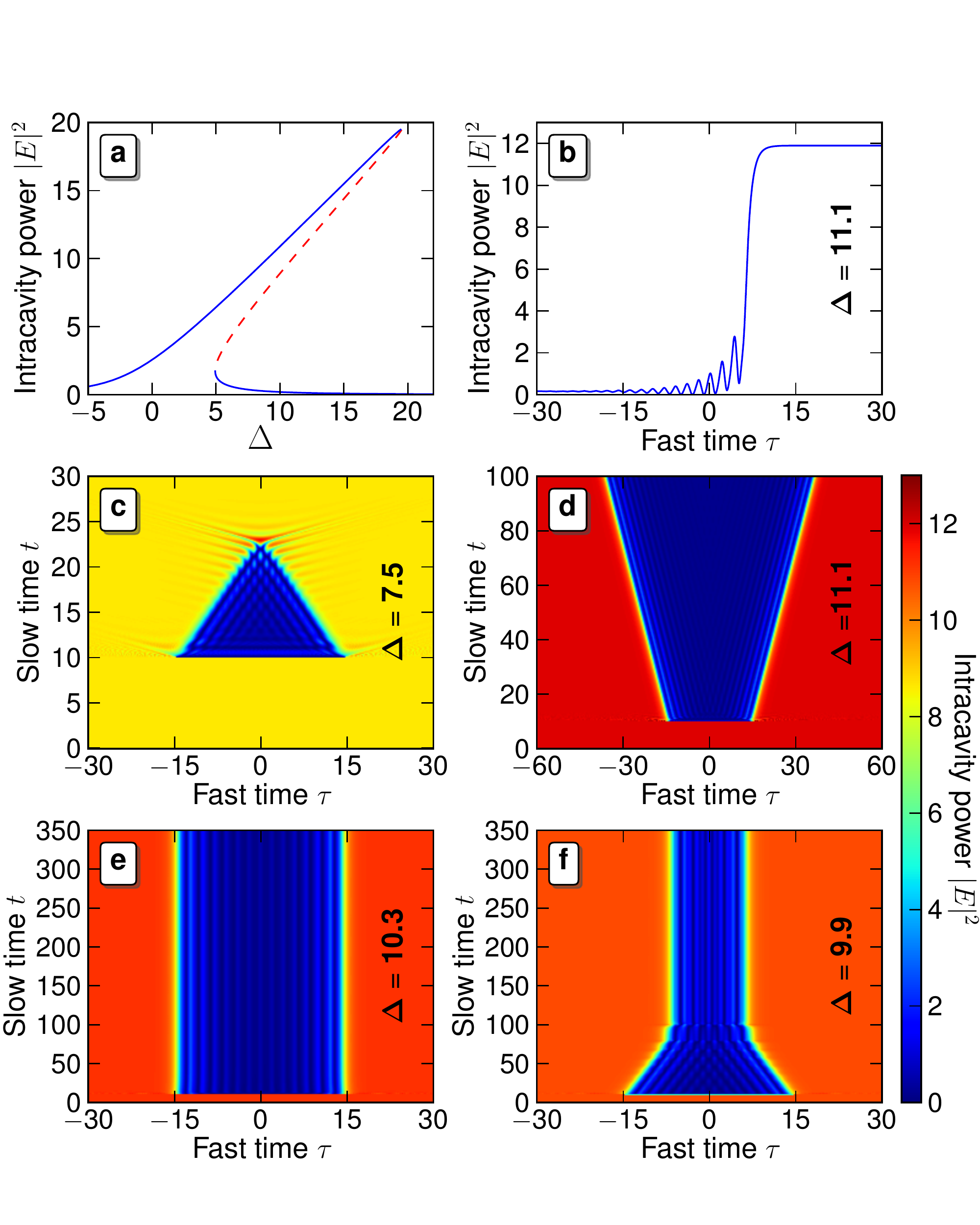}}
  \caption{Results of numerical simulation of the normalized LLE, obtained for a set driving power $X=19.5$. (a) Intracavity power of cw solutions as a function of cavity detuning $\Delta$. (b) Temporal power profile of a typical SW for $\Delta = 11.1$. (c)--(f) Pseudo-colour plots showing the evolution (bottom to top) of the intracavity power obtained for detunings $\Delta$ as shown in each panel.} \label{fig1}
\end{figure}

The homogeneous (cw) stationary solutions of Eq.~\eqref{LLE} satisfy the well-known cubic polynomial of dispersive optical bistability~\cite{haelterman_dissipative_1992}:
\begin{equation}
  X = Y^3-2\Delta Y^2+(\Delta^2+1)Y,
  \label{cubiceq}
\end{equation}
where $Y = |E|^2$ represents the intracavity power. For constant driving power $X$, these solutions describe a single Lorentzian cavity resonance that is tilted due to the Kerr nonlinear phase shift, and which can be expressed as
\begin{equation}
  \Delta = Y \pm \sqrt{\frac{X}{Y}-1}\,.
\end{equation}
Figure~\ref{fig1}(a) shows an example of such a tilted cw resonance, evaluated using parameters similar to the experiments that will be described in Section~\ref{expsection}. As is well known [and can be seen in Fig.~\ref{fig1}(a)], the cw solutions exhibit bistability for detunings between the up- and down-switching points $\Delta_\uparrow$ and $\Delta_\downarrow$, respectively (for the parameters of Fig.~\ref{fig1}(a), $\Delta_\uparrow = 4.9$ and $\Delta_\downarrow = 19.5$). Note that the intermediate branch [dashed red line in Fig.~\ref{fig1}(a)] is unconditionally unstable, and will not be discussed further.

In the regime of anomalous group-velocity dispersion ($\eta = -1$), the upper branch of the cw solutions exhibits MI over a wide region of parameters, and this process underpins the formation of Kerr optical frequency combs and TCSs~\cite{leo_temporal_2010,webb_experimental_2016}. In contrast, we are interested here in normally dispersive cavities ($\eta = +1$), where the two cw states are simultaneously stable across most of the bistable region (MI is limited to a small region around $\Delta_\uparrow$ at the end of the lower cw branch, and is not accessed in our experiments). In this case, both the lower and upper cw states can persist indefinitely (i.e.\ in steady-state), and coexist simultaneously at different positions along the resonator. Specifically, when operating in the bistable region, the temporally extended intracavity field can be composed of finite domains associated with the two different cw states. Because of the dispersive coupling, the domains are connected by fronts, or ``switching waves,'' featuring a smooth edge and an oscillatory tail~\cite{parra-rivas_origin_2016, oppo_domain_1999, ma_defect-mediated_2010}, as illustrated in Fig.~\ref{fig1}(b). In general, however, the cw states only co-exist for a limited time: the SWs move at a constant velocity, causing one type of domain to invade the other (one of the cw states can thus be considered as metastable~\cite{pomeau_front_1986}). The velocity of the SWs, as well as the relative stability of the two different cw states, depend on the control parameters \cite{parra-rivas_origin_2016, rosanov_diffractive_1990}. For fixed driving power~$X$, one can qualitatively appreciate that the lower (upper) cw state is more stable for detunings close to the down- (up-) switching point. It is only for one special set of parameters, known as the Maxwell point, that the two cw states are ``equally'' stable and can co-exist indefinitely in steady-state. For this set of parameters, the SWs exhibit of course zero velocity.

To illustrate this description, we show in Figs~\ref{fig1}(c)--(e) results from numerical simulations of the LLE that mimic the experiments that will follow. Here, we use a ``space-time'' representation that shows how the fast-time profile of the intracavity power (horizontal) evolves with slow-time (bottom to top), for a fixed driving power $X = 19.5$ and for a range of detuning values $\Delta$. In each case, the initial condition consists of the upper cw state, which is then perturbed --- at slow time $t$ equal to 10 --- by a super-Gaussian intensity dip that locally approximates the lower cw state. As seen in Fig.~\ref{fig1}(c), for a comparatively small $\Delta = 7.5$, the upper cw state quickly invades the lower one, while the opposite is true for a larger $\Delta = 11.1$ [Fig.~\ref{fig1}(d)]. For the driving power used in Fig.~\ref{fig1}, the Maxwell point occurs at $\Delta_\mathrm{M} \approx 10.3$, and indeed, our simulation reveals stationary SWs in this condition [Fig.~\ref{fig1}(e)].

Although the SWs are unconditionally stationary only at the Maxwell point, it is also possible for two SWs to ``lock'' in a finite range of parameters around the Maxwell point through interactions mediated by their oscillatory tails. Since oscillations are only present near the lower edge of the SWs [Fig.~\ref{fig1}(b)], this can only happen for a \emph{low} intensity domain embedded in a \emph{high} intensity background: the result is a dark localized structure (bright structures become possible e.g.\ in presence of high-order dispersion or nonlocality~\cite{parra-rivas_origin_2016}). In Fig.~\ref{fig1}(f), we illustrate such interlocking numerically with $\Delta = 9.9$ ($X = 19.5$ as before). As clearly seen, the two approaching SWs eventually come to a halt. It is precisely this interlocking behaviour that has recently been proposed to underlie the formation of mode-locked Kerr frequency combs in normally dispersive microresonators~\cite{parra-rivas_origin_2016, parra-rivas_dark_2016}.

For a given set of parameters, the SWs can lock for several discrete separations, corresponding to localized structures of different widths. Such multistability can be physically understood as locking between distinct cycles of the oscillatory SW tails, and is associated with a bifurcation structure referred to as \textit{collapsed snaking}~\cite{ma_defect-mediated_2010, parra-rivas_dark_2016}. Because the tail oscillations of the SWs are exponentially damped, one finds that structures with small widths generally exhibit larger regions of existence and stability~\cite{parra-rivas_origin_2016, rabbiosi_stochastic_2003}. Notwithstanding, the particular structures observed in simulations and experiments are not limited to those with the smallest widths, but rather are selected on basis of the initial condition (or the external perturbation) used to excite the SWs. For example, the final structure observed in Fig.~\ref{fig1}(f) corresponds to the largest stable separation for two SWs for the chosen parameter values (here the SW separation $\Delta\tau=12$ in normalized units).

\section{Experimental setup \label{expsection}}

\begin{figure}[t]
  \centering
  \includegraphics[width=1.0 \columnwidth]{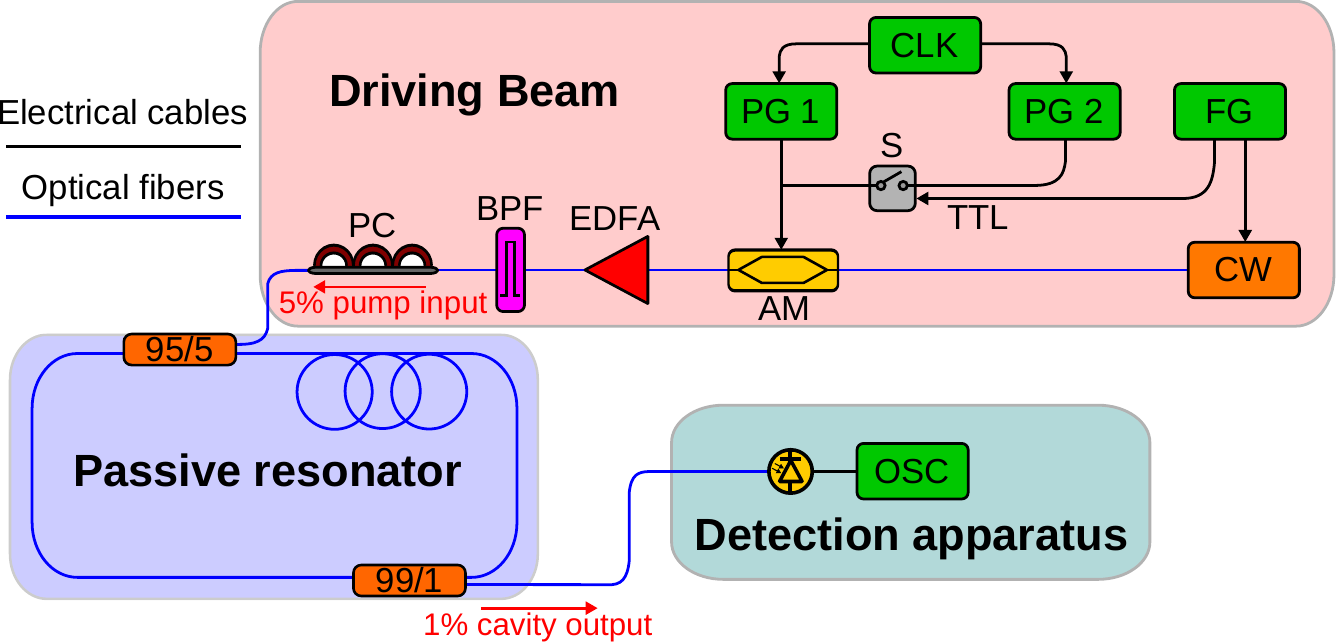}
  \caption{Schematic illustration of the experimental set-up. cw, tunable cw driving laser; FG, function generator; PG 1 and PG 2, pattern generators as described in the text; CLK, clock; S, electrical switch; AM, Mach-Zehnder intensity modulator; EDFA, Erbium-doped fibre amplifier; BPF, band-pass filter; PC, polarization controller; OSC, oscilloscope.}\label{fig2}
\end{figure}

In this work, our aim is to experimentally investigate the temporal SW dynamics described above. To this end, we use the setup illustrated in Fig.~\ref{fig2}. It consists of a passive fibre ring resonator with a total length of about 90~metres. The cavity is mainly constructed of a dispersion shifted fibre (Corning MetroCor, 88~metres long) that exhibits normal dispersion ($\beta_2 = 8.1~\mathrm{ps^2/km}$) at the driving wavelength of 1550~nm, but also includes about $0.7$~m of standard SMF-28 fibre with anomalous dispersion ($\beta_2 = -21.4~\mathrm{ps^2/km}$). This yields a round-trip time $t_\mathrm{R}$ of about {430~ns}, corresponding to a free-spectral-range (FSR) of $2.3$~MHz, and a cavity averaged dispersion of $\langle\beta_2\rangle = 7.9~\mathrm{ps^2/km}$. The resonator also includes a 95/5 coupler used to inject the driving field into the cavity, as well as a 99/1 coupler through which the intracavity dynamics are monitored using a $12.5$~GHz photodetector connected to a real time oscilloscope. Overall, the cavity has a high measured finesse of~37, corresponding to a photon lifetime of about 6~round trips. 

To drive the cavity, we use flat-top, quasi-cw, nanosecond pulses at a repetition rate synchronized to the cavity FSR~\cite{coen_modulational_1997, copie_competing_2016, anderson_observations_2016, anderson_super_2017}. These pulses are obtained by intensity modulating a commercial, wavelength-tunable, Erbium-doped distributed feedback cw fibre laser (Koheras AdjustiK{\scriptsize\texttrademark} E15) by means of a Mach-Zehnder modulator (JDSU, $10~\mathrm{Gb/s}$ intensity modulator) driven by a 10~GHz pattern generator (PG~1 in Fig.~\ref{fig2}). After the modulator, the pulses are amplified up to $1.3$~W peak power (corresponding to a normalized driving power $X = 19.5$) using an Erbium-doped fibre amplifier (Keopsys 5~W), and spectrally filtered to remove amplified spontaneous emission. A polarization controller is finally used to set the state of polarization of the driving pulses before their injection into the cavity.

\begin{figure}[t]
  \centerline{\includegraphics[width=1.0 \columnwidth]{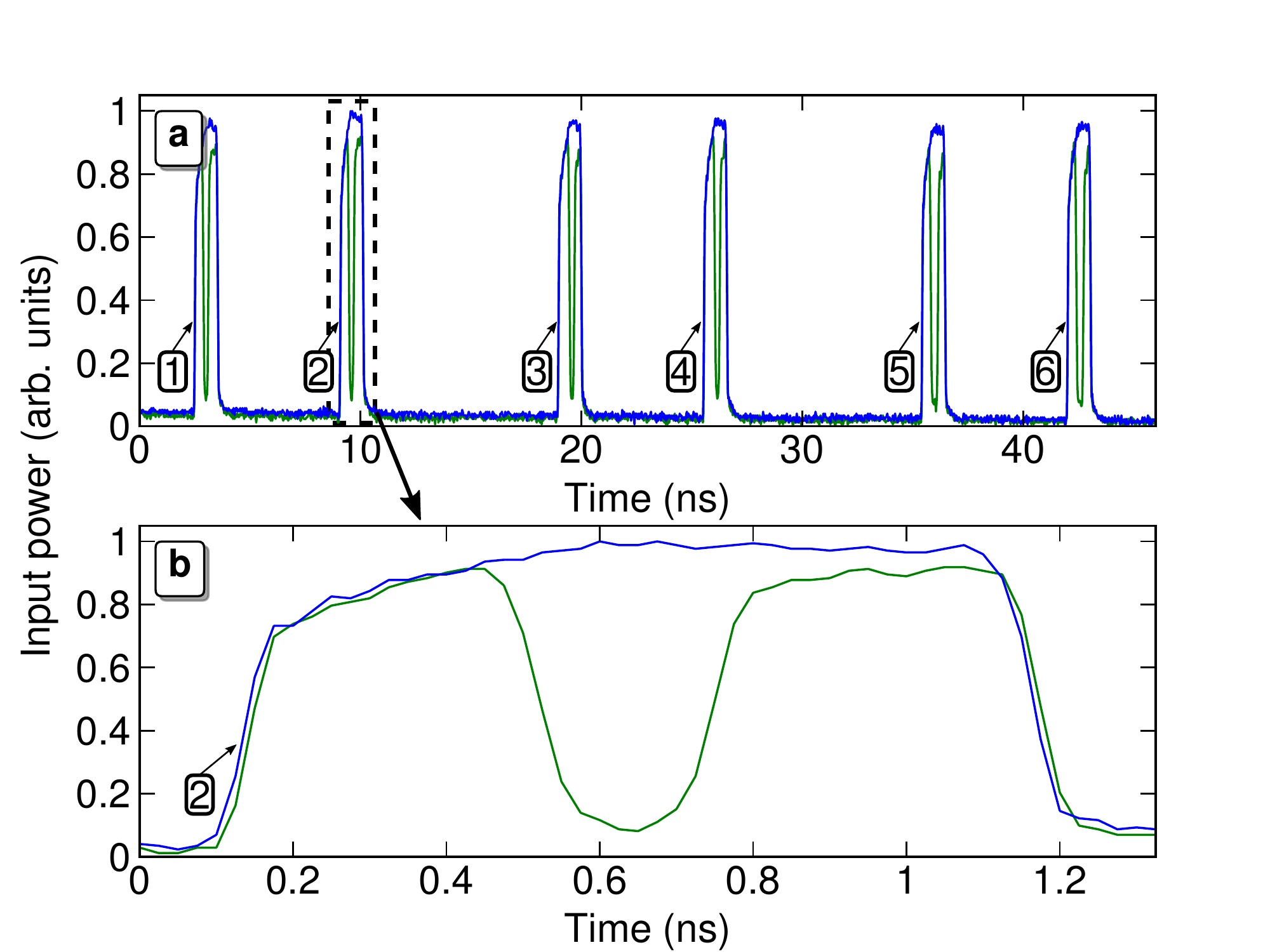}}
  \caption{(a) Pattern of quasi-cw pulses used to drive the resonator. The pattern repeats every 430~ns, matching the cavity round-trip time, thus ensuring synchronous driving. Each pulse is labelled numerically to facilitate the discussion of our results (the labels match those in Fig.~\ref{fig5}). (b) Enlarged view of driving pulse~\#2. Blue (green) curves show the driving pulses in the absence (presence) of the perturbations used to excite SWs.}
  \label{fig3}
\end{figure}

The reason for driving the cavity with nanosecond pulses (instead of a purely cw field) is threefold. Firstly, pulses can be readily amplified to larger peak powers than cw fields, allowing access to larger resonance tilts and therefore larger detunings~\cite{anderson_observations_2016, anderson_super_2017}. Secondly, the large bandwidth of the driving pulses suppresses the detrimental effects of stimulated Brillouin scattering (SBS). We note that previous fibre resonator studies involving TCSs in the anomalous dispersion regime relied on an intracavity isolator to mitigate SBS~\cite{leo_temporal_2010, jang_ultraweak_2013, jang_temporal_2015}. This method is not applicable to our experiment because, in contrast to TCSs which always sit on a \emph{lower} state background, parts of our intracavity field are in the \emph{upper} cw state, with an intensity level much more susceptible to SBS. On the plus side, the lack of an isolator enables us to reach comparatively higher cavity finesses. Thirdly,  by driving the cavity with a pulse pattern consisting of \emph{multiple} widely-spaced nanosecond pulses per round trip (the whole pattern being synchronized to the cavity FSR), we can simultaneously realize several independent experimental realizations. The driving pattern used in our experiment contains six nanosecond-duration pulses and is illustrated in Fig.~\ref{fig3}(a) [blue curve]. As can be seen, small differences in peak power levels exist between the different driving pulses due to modulation and amplification imperfections. This causes small artefacts in our observations which will be discussed below.

Because the nanosecond driving pulses are much longer than the picosecond-scale SWs we are interested in, the driving field can be considered as quasi-cw. We note, however, that a small synchronization mismatch between the repetition rate of the driving pattern and the cavity FSR is generally unavoidable and results in an effective convective drift~\cite{coen_convection_1999} that can hinder the interpretation of the data. In particular, the front expansion/contraction dynamics can appear asymmetric if examined in a (natural) reference frame fixed to the driving pattern repetition rate. To alleviate this issue, all our results are presented in a reference frame fixed to the cavity FSR, in which the SW motion is symmetric. (Of course, in this reference frame, a synchronization mismatch causes the nanosecond background to exhibit a uniform temporal drift.)

Finally, to systematically excite SWs, we use a signal from a second pattern generator (PG~2 in Fig.~\ref{fig2}) synchronized to the driving pattern to perturb selected driving pulses. That signal is made up of approximately Gaussian pulses, shorter than the driving pulses, and generated with a $\pi$ phase shift with respect to that of PG~1. When fed through the electrical switch (S), it thus effectively carves dips in the nanosecond driving pulses. The green curve in Fig.~\ref{fig3}(a) illustrates the resulting perturbed driving pulse pattern [with an enlargement of one the pulses in Fig.~\ref{fig3}(b)] compared to the unperturbed driving (blue curve). Note that the switch is only open during the initial phase of SW excitation and is blocked for the rest of the experiment. This effectively mimics the procedure used to excite SWs in the simulations presented in Fig.~\ref{fig1}.

%In the absence of tail's oscillations (i.e. monotonic relaxation), this unique point is expected to exhibit stationary SWs propagation. However, precisely accessing (and staying at) the Maxwell point is challenging in our current experiments, as (i) the point corresponds to a unique set of parameters and (ii) we do not employ any active stabilization scheme to lock the cavity detuning (or the driving strength). We nevertheless remark that a simple stabilization scheme that locks the average intracavity power to a set level automatically locks the system at the Maxwell point, as any SW motion would at any other point change the average power.

\section{Experimental results}

\begin{figure}[b]
  \centerline{\includegraphics[width=1.0 \columnwidth]{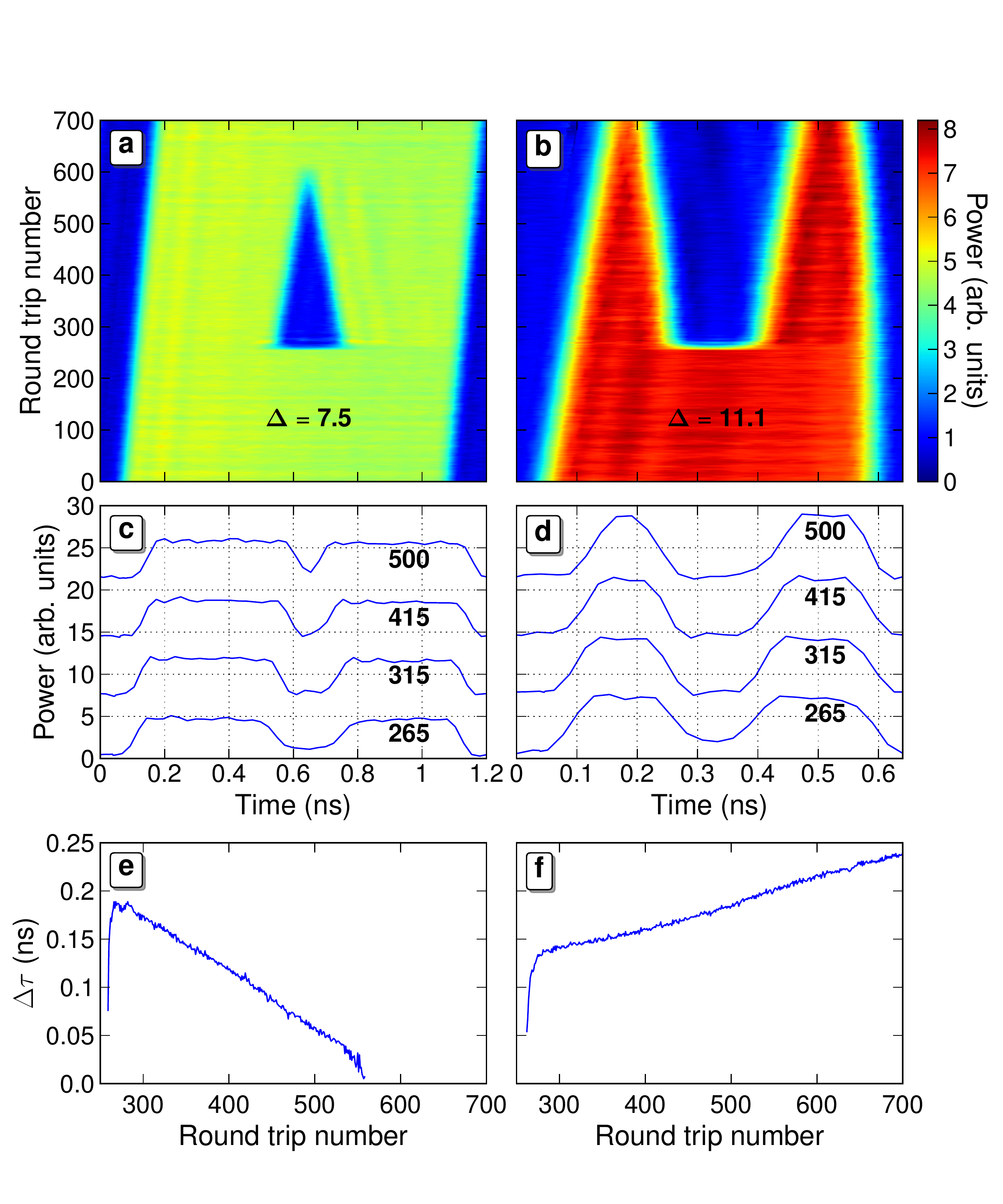}}
  \caption{Experimental measurements of SW dynamics, obtained for two different values of the cavity detuning as indicated. (a, b) Space-time diagrams showing the evolutions of the temporal profile of the cavity output power over 700 cavity round trips. The SWs are excited starting from round trip~\#253. (c, d) Snapshots of the cavity output power profile at specified round trips, after excitation of the SWs. (e, f) Corresponding evolution of the temporal separation of the two SWs, $\Delta\tau$.}\label{fig4}
\end{figure}

To investigate SW dynamics in a systematic way, we start each measurement by tuning the frequency of the driving laser with a function generator in order to select a detuning value $\Delta$ that lies in the regime of cw bistability. Specifically, we ramp up the detuning across the cavity resonance and onto the upper branch, i.e.\ from left to right in Fig.~\ref{fig1}(a), and stop at a selected point of interest. In this way, the intracavity field initially corresponds to the upper cw state. Next, we perturb the driving pulses for about 10~cavity round trips (with the S~switch, as described above), which locally switches the intracavity field to the lower cw state, creating SWs on both sides. After the switch is returned to its original off position, and the perturbation has died off, the driving pulses regain their original flat-top shape (blue curve in Fig.~\ref{fig3}) and the SWs are left to evolve unperturbed. The excitation and evolution of the SWs are recorded as a long time trace of the cavity output power on an oscilloscope.

Typical experimental results, obtained for two different cavity detunings, are shown in Figs~\ref{fig4}(a) and~(b). Here we visualize the intracavity field evolution by means of the same space-time representation as used in Figs~\ref{fig1}(c)--(f), and that is obtained by (i) slicing a single long oscilloscope trace into segments separated by the cavity round-trip time and (ii) concatenating the individual segments atop each other. Note that, as explained above, the visible drift of the nanosecond background pulse originates from a mismatch between the repetition rate of the driving pattern and the cavity FSR. As can be seen, for a small detuning $\Delta = 7.5$ [Fig.~\ref{fig4}(a)], the domain of lower cw state (induced by the applied perturbation) contracts at the expense of the surrounding upper state, with the two SWs moving towards each other. In contrast, for a larger detuning $\Delta = 11.1$, the lower state domain expands, with the SWs moving away from each other [Fig.~\ref{fig4}(b)]. In Figs~\ref{fig4}(c) and~(d), we show corresponding snapshots of the oscilloscope traces at four different round trips, as indicated in the panels.

The results shown in Figs~\ref{fig4}(a)--(d) clearly illustrate the motion of SWs. To obtain more quantitative insights, we show in Figs~\ref{fig4}(e) and~(f) the evolution of the temporal separation between the SWs, $\Delta\tau$. That quantity represents the temporal duration of the lower cw state domain and is extracted from our data as $\Delta\tau = \tau_2-\tau_1$, where $\tau_2$ ($\tau_1$) is the temporal position of the trailing (leading) SW. We can clearly see that the domain size changes linearly over time, i.e.\ the SW moves at constant velocity. By extracting the slope of the curve, we can readily obtain a quantity that characterizes the SW velocity,
\begin{equation}
V = \frac{1}{2}\frac{\mathrm{d}\Delta\tau}{\mathrm{d}t/t_\mathrm{R}},
\end{equation}
where the factor of two accounts for the symmetric motion of the two SWs. For the measurements in Figs~\ref{fig4}(a,c) and~(b,d), we obtain $V \approx -0.29$~ps per round trip and $V \approx +0.13$~ps per round trip, respectively. Note that, with our definitions, $V<0$ ($V>0$) implies contraction (expansion) of the lower cw state, i.e.\ the SWs moving towards (away from) each other.

\begin{figure}[b]
  \centerline{\includegraphics[width=1.0 \columnwidth]{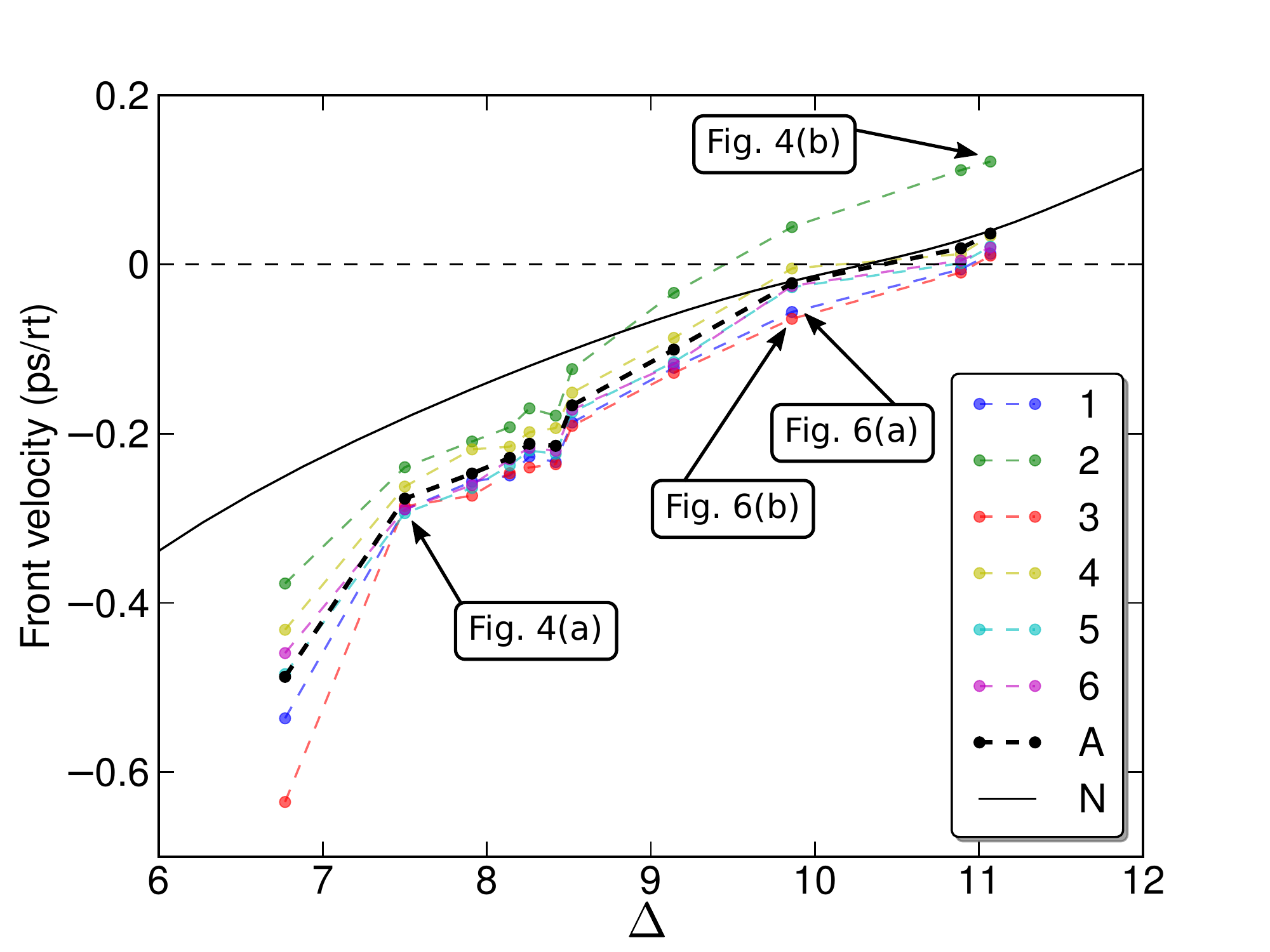}}
  \caption{Experimentally measured SW velocities as a function of cavity detuning $\Delta$. The different numerically-labelled curves correspond to the different driving pulses shown in Fig.~\ref{fig3}(a). The black dashed curve (A) is the average over all realizations while the solid black curve (N) is the numerical prediction. Points corresponding to data displayed in other figures are labelled accordingly.}
  \label{fig5}
\end{figure}

We have repeated the measurements over a wide range of cavity detunings, and summarize our findings in Fig.~\ref{fig5}. Here we plot the measured SW velocities as a function of the detuning, with the different curves corresponding to the different pulses of the driving pattern (coloured curves with labels coinciding with those in Fig.~\ref{fig3}). We believe that slight variations in the peak power levels of the different driving pulses (see Section~\ref{expsection}) explain the small spread of SW velocities observed in Fig.~\ref{fig5}. To facilitate our analysis, we also plot in Fig.~\ref{fig5} the average SW velocity computed across all the different driving pulses (dashed black curve, labelled `A').

%We note that, \textcolor{red}{because of why oh why}, our measurements are limited to small and intermediate values of the cavity detuning (compared to the down-switching point $\Delta_\downarrow = 19.5$), with a maximum accessed value of $\Delta_\mathrm{max} \approx 11.1$. However, as described below, the range of detunings accessed in our experiments is sufficient for the observation of all dynamical regimes of interest.

\begin{figure}[b]
  \centerline{\includegraphics[width=1.0 \columnwidth]{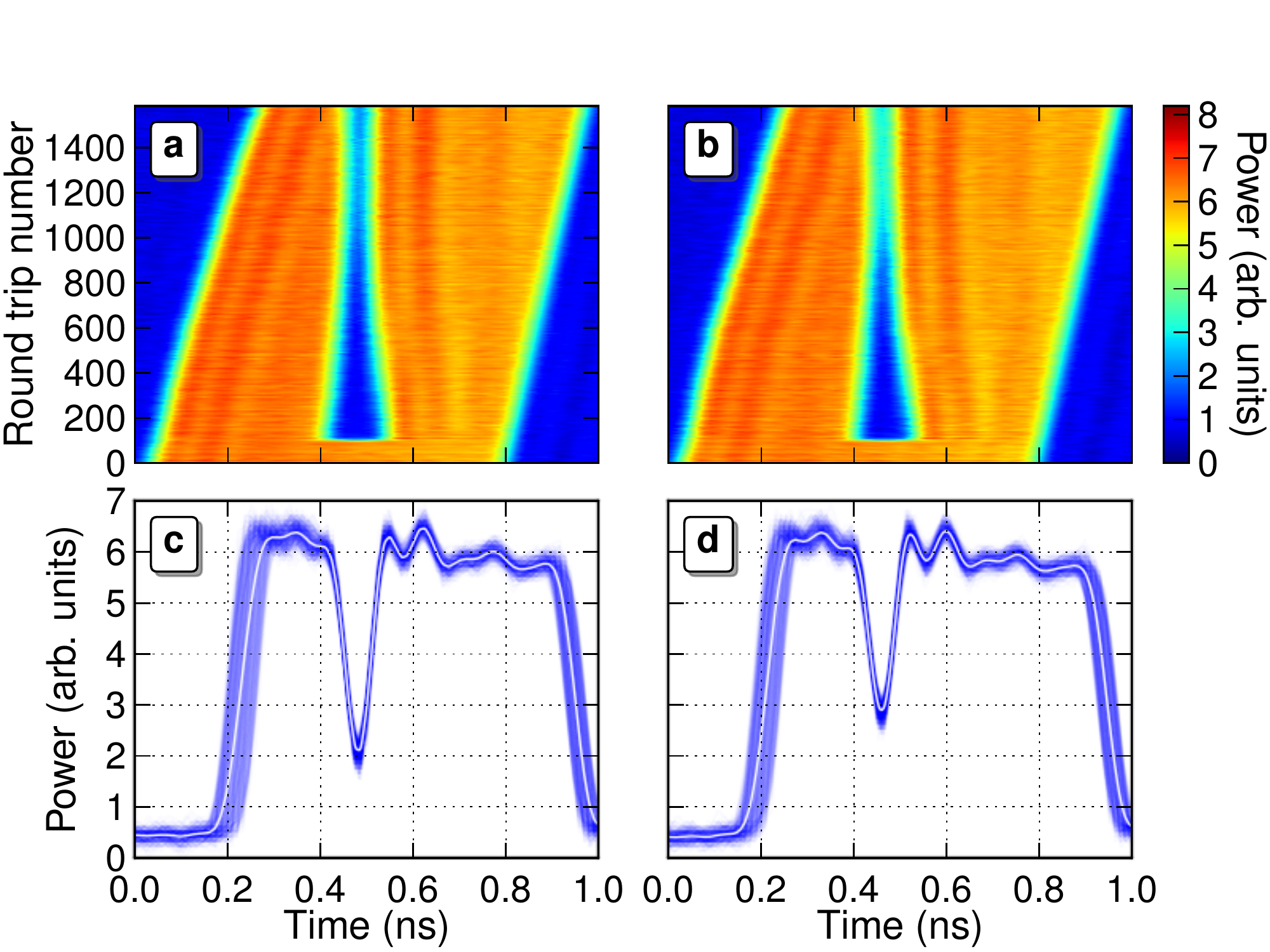}}
  \caption{Experimental measurements showing evidence of SW interlocking, obtained for $\Delta \approx 9.9$. (a, b) Space-time diagrams of the cavity output power corresponding, respectively, to driving pulses \#1 and~\#3 as labelled in Fig.~\ref{fig3}(a). (c, d) Superpositions of oscilloscope traces corresponding to the last 400 round trips of the data shown in (a, b). The white lines represent the average of the traces displayed.}
  \label{fig6}
\end{figure}

The experimental results in Fig.~\ref{fig5} clearly show that, for small detunings, the lower cw state is metastable: it is quickly invaded by the upper state, with the SWs moving towards each other [$V<0$, as in Fig.~\ref{fig4}(a)]. As the detuning increases, the SW speed $V$ decreases, and crosses zero at about $\Delta_\mathrm{M} \approx 10.4$ (average value for the six independent realizations). For larger detunings, we see clearly how it is now the upper cw state that is metastable, shrinking at the expense of the lower state (i.e.\ $V>0$). The zero-crossing of the SW velocity corresponds, of course, to the Maxwell point for our parameters. In this context, we highlight that the value $\Delta_\mathrm{M}\approx 10.4$ extracted from our measurements is in very good agreement with the value obtained from numerical simulations [$\Delta_\mathrm{M}\approx 10.3$, see Fig.~\ref{fig1}(e)]. Numerically calculated SW velocities are also plotted in Fig.~\ref{fig5} (solid black curve, labelled `N') for the full range of detunings considered in the experiment and are found to match the measurements reasonably well.

%In the absence of tail's oscillations (i.e. monotonic relaxation), this unique point is expected to exhibit stationary SWs propagation. However, precisely accessing (and staying at) the Maxwell point is challenging in our current experiments, as (i) the point corresponds to a unique set of parameters and (ii) we do not employ any active stabilization scheme to lock the cavity detuning (or the driving strength). We nevertheless remark that a simple stabilization scheme that locks the average intracavity power to a set level automatically locks the system at the Maxwell point, as any SW motion would at any other point change the average power.

The SW velocities shown in Fig.~\ref{fig5} were extracted for SWs evolving independently, i.e.\ without interaction. Of course, departure from purely linear front motion can be expected when the SWs get sufficiently close to each other for their oscillating tails to overlap~\cite{parra-rivas_origin_2016}. To consider this situation, and the possible interlocking of SWs, we have analyzed data acquired for a detuning $\Delta\approx 9.9$, just below the Maxwell point, over a larger number of round-trips. These measurements are presented in Figs~\ref{fig6}(a) and~(b), again in the form of space-time diagrams. Data shown in these two panels correspond to two different driving pulses [respectively \#1 and~\#3 as labelled in Fig.~\ref{fig3}(a)] but were otherwise acquired simultaneously, i.e.\ for the exact same detuning. Only the driving powers are slightly different as already mentioned. As can be seen, the lower state domain initially contracts, as in Fig.~\ref{fig4}(a), but then, instead of merging together, the two SWs slow down and eventually come to a halt, giving rise to a persisting localized dark structure.
\begin{figure}[t]
  \centerline{\includegraphics[width=1.0 \columnwidth]{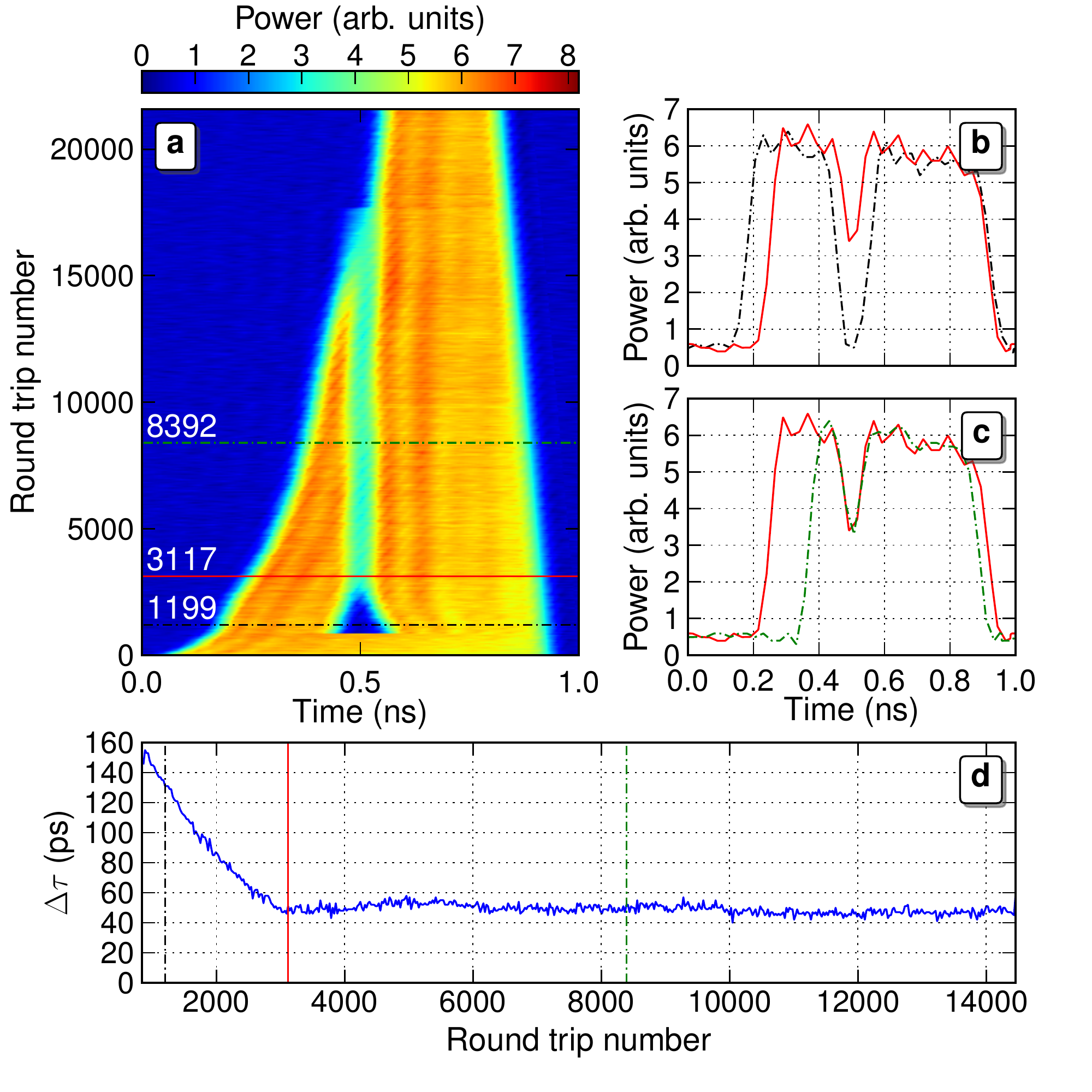}}
  \caption{Experimental results showing the long-term stability of interlocked SWs. The data was obtained for $\Delta \approx 9.4$ with driving pulse \#2. (a) Space-time diagram of the evolution of the temporal power profile at the cavity output. (b, c) Selected snapshots; the colours correspond to the horizontal lines in (a), with the corresponding round trip number as indicated. (d) Corresponding evolution of the separation of the two SWs, $\Delta \tau$.}
  \label{fig7}
\end{figure}

The temporal intensity profiles of the observed dark structures are shown in more detail in Figs~\ref{fig6}(c) and~(d), respectively for each of the two driving pulses considered here. In this plot, we overlay oscilloscope traces corresponding to the last 400 round trips of our measurements, highlighting how the observed structures maintain a constant shape and power level. Unfortunately, the limited 12~GHz bandwidth of our detection system is insufficient to properly resolve the optical dark structures, which according to simulations should not be larger than 25~ps. In this context, we suspect that the difference in the depth of the dark dips registered in Figs~\ref{fig6}(c) and~(d) is an artefact, and is rather indicative of dark structures with different temporal durations. Indeed, for timescales shorter than the impulse response, the detector simply yields a measurement proportional to the signal \emph{energy}. As shorter dips lead to a reduced drop in detected energy, one expects them to register as a shallower feature on the oscilloscope.

The results shown in Fig.~\ref{fig6} are indicative of SW interlocking. However, because of the limited memory depth of our oscilloscope, the measurement only encompasses about 1000 round trips, preventing full conclusions on the long term stability of the dark structures. To gain more convincing evidence, we performed another set of measurements for which we captured the cavity output only once every 23 round trips, leading to a coarser but significantly longer acquisition. The result is presented in Fig.~\ref{fig7}. As in Figs~\ref{fig6}(a) and (b), we observe an initially shrinking lower state domain, and the associated SWs eventually interlock around the 3000th round trip. This behaviour can be observed in more details with the selected snapshots plotted in Figs~\ref{fig7}(b) and (c), and also in the evolution of the separation between the two SWs [Fig.~\ref{fig7}(d)], which presents a stage of linear decrease followed by a characteristic plateau. The formed dark structure is seen to persist unchanged for more than 10000 round trips, which corresponds to almost 2000 photon lifetimes. Eventually, around round trip \#17500, the dark structure is destroyed as it collides with another SW that has emerged spontaneously from the leading edge of the driving pulse. Nevertheless, the fact that the dark structure persists for almost 2000 photon lifetimes is clear evidence of stable SW interlocking.

\section{Conclusions}

In conclusion, we have experimentally and theoretically investigated the dynamics of SWs in a passive Kerr fibre ring resonator operating in the normal dispersion regime. By locally perturbing the homogeneous intracavity field in a region of cw bistability, we were able to systematically generate SWs and investigate their dynamics. In particular, we measured the velocities of the SWs as a function of the cavity detuning, observing a zero crossing at the point of equal stability (the Maxwell point). Additionally, for parameters sufficiently close to the Maxwell point, we observed clear evidence of interlocking of two SWs and the corresponding formation of persisting dark temporal structures. Evidence for the multistability of the structure width have also been presented, but are limited by the resolution of our detection system. All our experimental results are in good agreement with simulations of the Lugiato-Lefever equation. Our findings help link recent theoretical and numerical works on SWs~\cite{parra-rivas_origin_2016} to the generation of dark pulse Kerr frequency combs observed in normally dispersive microresonator experiments~\cite{xue_mode-locked_2015}, and provide insights on the operation of coherently driven fibre ring resonator in the normal dispersion regime.

\section*{Acknowledgements}

We gratefully acknowledge support from the Marsden Fund and the Rutherford Discovery Fellowships of the Royal Society of New Zealand.

\newcommand{\singleletter}[1]{#1}

\bibliographystyle{bibstyle2nonotes}
\bibliography{stephane,microcavity,sw,cs}

\end{document}